\documentclass[prl,twocolumn,floatfix,nobibnotes,superscriptaddress,prb]{revtex4}
\usepackage{amssymb}
\usepackage{amsmath}
\usepackage[dvips]{graphicx}
\usepackage[usenames,dvipsnames]{color}
\usepackage[english]{babel}
\usepackage[latin1]{inputenc}
\usepackage{latexsym}

\usepackage{dcolumn}
\usepackage{bm}
\usepackage{multirow}

\begin{document}
\newcommand{\red}{\color{black}}
\newcommand{\blue}{\color{black}}
\title{Evolving properties of two dimensional materials, from graphene to graphite}

\author{Mattias Klintenberg}
 \email[Corresponding author: ]{Mattias.Klintenberg@fysik.uu.se}
 \affiliation{Department of Physics and Materials Science, Uppsala University, Box 530, 751 21 Uppsala, Sweden}
 \author{{\red S\'ebastien Leb\`egue}}
 \affiliation{
 Laboratoire de Cristallographie, R\'esonance Magn\'etique et Mod\'elisations (CRM2, UMR CNRS 7036)
 Institut Jean Barriol, Nancy Universit\'e
 BP 239, Boulevard des Aiguillettes
 54506 Vandoeuvre-l\`es-Nancy,France 
 }
\author{{\red Carlos Ortiz}}
 \affiliation{Department of Physics and Materials Science, Uppsala University, Box 530, 751 21 Uppsala, Sweden}
 \author{Biplab Sanyal}
 \affiliation{Department of Physics and Materials Science, Uppsala University, Box 530, 751 21 Uppsala, Sweden}
\author{Olle Eriksson}
 \affiliation{Department of Physics and Materials Science, Uppsala University, Box 530, 751 21 Uppsala, Sweden}

\date{\today}

\begin{abstract}
We have studied theoretically, using density functional theory, several materials properties when going from one C layer in graphene to two and three graphene layers and on to graphite. The properties we have focused on are the elastic constants, electronic structure (energy bands and density of states), and the dielectric properties. For any of the properties we have investigated the modification due to an increase in the number of graphene layers is within a few percent. Our results are in agreement with the analysis presented recently by Kopelevich and Esquinazi (unpublished).

\end{abstract}

\maketitle

\section{\label{introduction}Introduction}
The recent explosion of scientific activity around the newly discovered two-dimensional material, graphene, is unprecedented since the discovery of the high temperature superconductors in the late 1980'ies.\cite{ref1,ref2,ref3,ref4,ref5}. The uniqueness of this material, and the technological advantages it promises, gathers researchers from different scientific fields. As in many previous major scientific breakthroughs the main component for the success story of graphene was the actual synthesis of the material\cite{novoselov}. Unexpectedly Novoselov et al. were able to fabricate a truly two-dimensional material, and free standing single graphene layers, the building block of graphite, produced by means of exfoliation. Graphene can also be grown on SiC by epitaxial growth. In addition growth of graphene on catalytic surfaces (e.g. Ni or Pt) has been demonstrated. An insulating thicker material can be grown on top and after chemically removing the primary layer, one is left with a single atomic layer of graphene on an insulating substrate\cite{ref1}.

Research on graphene was initially motivated by a highly spectacular phenomenon, namely mass-less Dirac fermions. Although there is nothing relativistic in the single electron Kohn-Sham Hamiltonian describing the electronic structure of graphene, the band dispersion turns out to have a very unique property, at least from calculations based on the local density approximation (LDA) and the generalized gradient approximation (GGA), which enables a comparison to quasiparticle states obtained from the Dirac equation, with an effective speed of light of ~10$^6$ m/s \cite{ref1,ref3} and zero rest mass. 

Studies of the electronic properties of graphene have revealed an ambipolar electric field effect, with high concentrations and high mobility (up to 15.000 cm$^2$/Vs). In addition, conductivity properties reveal ballistic transport on the submicrometer scale, which is unexpected not least due to that the measurements are not done in ultra high vacuum, and hence many different molecular species are expected to be absorbed and act as scattering centers. Furthermore, graphene is the only material know to date with a quantum Hall effect (QHE) at room temperature\cite{QHE}, and the QHE was shown to be somewhat anomalous in nature\cite{QHE}. The realization of a material with negative index of refraction for electrons has also been demonstrated in graphene, which is due to that the electron states in the valence band have a group velocity antiparallel to the k-vector.  Hence, the Veselago lens, which has the unique property of having a resolution not determined by the wavelength, has been demonstrated in applications with graphene\cite{veselago}.

The focus of the present paper is to investigate how the electronic properties of graphene evolves to those of graphite, by a systematic theoretical study of 1, 2 and 3 layers of graphene. We will make comparisons between the calculated results and existing data for graphite, and in this way we will shine light on how the electronic structure of sp$^2$ bonded C layers evolve from that of graphene to that of graphite. The properties we focus on here are the electronic structure, the dielectric function, $\epsilon(\omega)$, and the elastic constants.

\section{\label{details}Details of Calculations}
{\red The electronic structure and elastic constats have been calculated using a highly accurate full potential linear muffin-tin orbital (FP-LMTO) method \cite{wills} within the local density approximation (LDA). A  $36 \times 36 \times 6$ k-point grid gave converged results. The FP-LMTO and PAW (see below) results are found to agree nicely. A $2sp3sp$ basis set was necessary to obtain a correct electronic structure, {\it c.f.} the minimal $2sp$ basis set, and the muffin-tin radii were optimized to cover 90\% of the nearest neigbour distance.}

The calculations of the optical properties have been performed by using the VASP (Vienna Ab-initio Simulation Package) code\cite{VASP1,VASP2},
implementing the PAW formalism\cite{PAW}. We have used the PBE exchange-correlation functional\cite{PBE}. To insure negligible interaction
  between periodic images, a large value ($20$ \AA) of the cell parameter 'c' was used. The convergence of the dielectric
  function is obtained by using a $80 \times 80 \times 5$ Monkhorst-Pack mesh\cite{Monkhorst:1976}. For the plane wave expansion of the wave function a $400$ eV cut-off was used.

\section{\label{results}Results}
\subsection{Electronic structure}
The energy bands for 1, 2 and 3 graphene layers are shown in Fig.1. For one graphene layer we find an electronic structure which is similar to 
 that found by others, for instance in Ref.{\red \cite{wallace,mcclure,reich}}. At the K-point two energy bands of p$_z$ character cut the Fermi level (E$_F$), and the energy 
 dispersion is (close to) linear with respect to the crystal momentum. This represents the bands refered to as mass-less Dirac Fermion states.
 For two layers the number of energy bands doubles, and there are four sets of p$_z$ derived bands close to the K-point. Due to the interaction
 between the graphene layers these bands split apart so that only two bands cut E$_F$, and the energy dispersion deviates more from linear 
 compared to the situation for one layer. For three layers a set of six p$_z$ derived bands can be found close to E$_F$ at the K-point. None of the bands cut E$_F$, i.e. a small gap is introduced between four of these bands, whereas two are split, and are found further away either above or below E$_F$. In general Fig.1 suggests that the more carbon 
 layers that are introduced the wider energy range does the set of p$_z$ bands span. This saturates for bulk graphite where in addition to 
 the degenerate bands at E$_F$ there is one band $\sim$ 0.7 eV above E$_F$ and one band $\sim$ 0.7 eV below E$_F$.\cite{rajeev}

The corresponding density of states are shown in Fig. \ref{fig2}.

The three layers bandstructure look like a combination of the bandstructure of one graphene layer
 and two layers with the exeption that a small bad gap has been introduced ($0.01$ eV). We also note that the FP-LMTO results presented here do not show the band-overlaps around the K-point at E$_F$ as do the tight-binding results in.\cite{partoens} 

\begin{figure*}[htbp]
\includegraphics*[angle=0,scale=0.50]{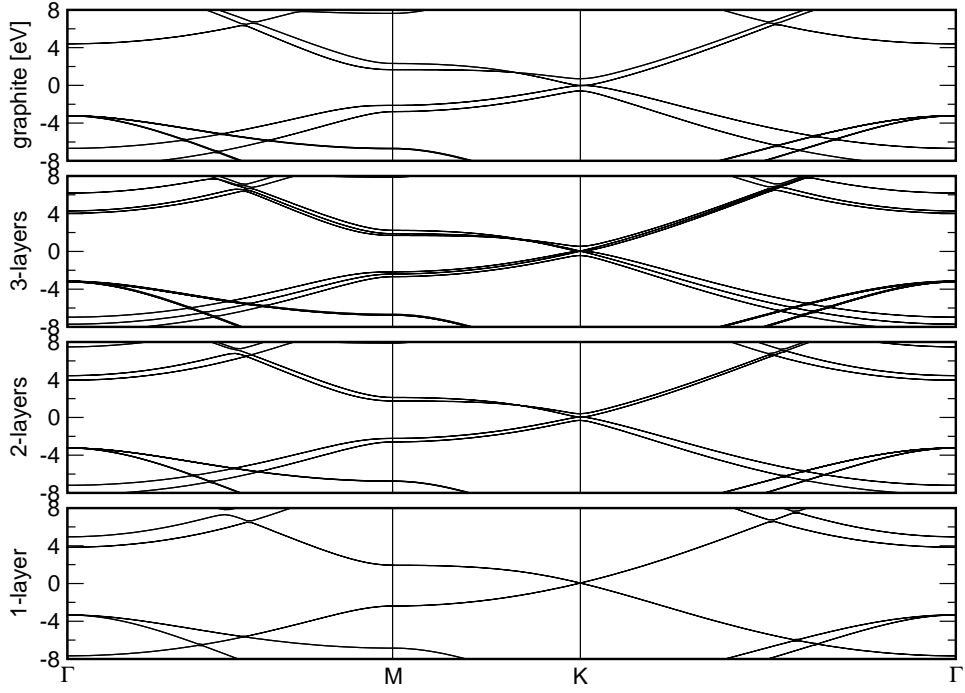}
\caption{{\red The LDA electronic bandstructure for 1-, 2-, 3-layers graphene and graphite. The linear bands around the K-point near the fermi level can be seen for 1-, and 3-layers, {\it i.e.} odd number of layers.}}
\label{fig1}
\end{figure*}

\begin{figure*}[htbp]
\includegraphics*[angle=0,scale=0.50]{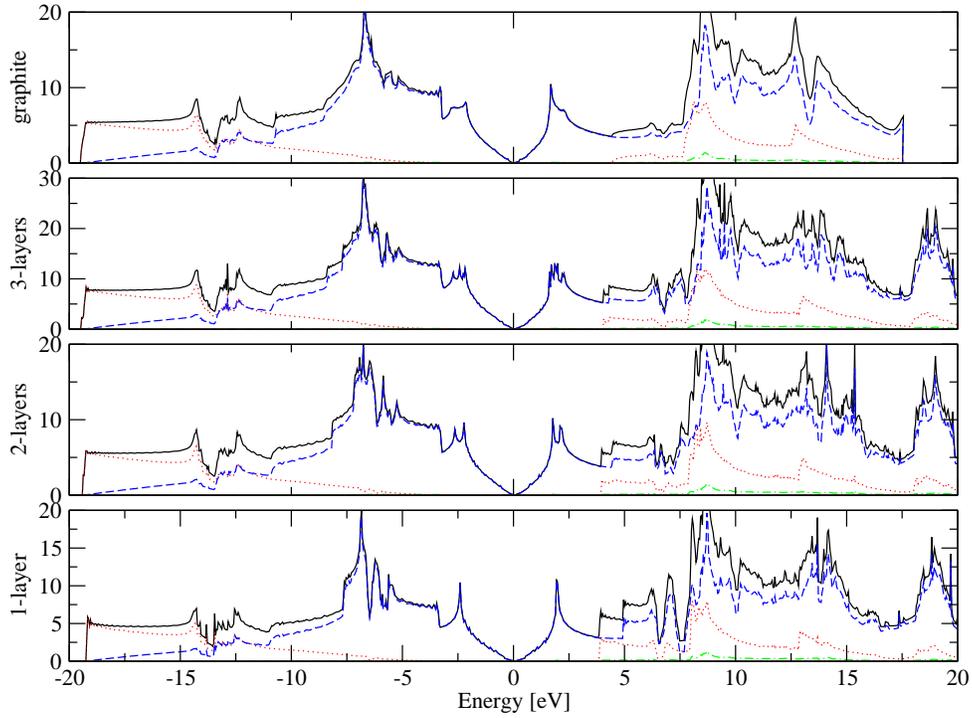}
\caption{{\red The density of states for 1-, 2-, 3-layers graphene and graphite. Colored lines show $l$-resolved partial DOS, $s$ (dot,red), $p$ (dash,blue), and $d$ (dot-dash,green). Black full line shows Total DOS.}}
\label{fig2}
\end{figure*}

Some experiments, using either the de Haas van Alphen effect\cite{dHvA}, or the quantum Hall
effect\cite{Hall}, have shown that electrons and holes with linear dispersion relations could exist
  not only in graphene but also in graphite. In fact, it was shown using a tight-binding 
   description of the electronic structure\cite{tightb} that Dirac fermions are existing in graphene
    multilayers if the number of layers is odd (systems with a mirror inversion plane).
However, tight-binding calculations depend on a given set of parameters, and although they can provide clues about the 
 general mechanisms, they are not as precise as ab-initio calculations. Therefore it was important to check
  if the above assumptions are still valid when obtained from ab-initio calculations. 
  We computed precisely the electronic structure of each system around the $K$ point, as shown in Fig.1
  and found that indeed linear bands are present for one and three layers of graphene, but not for two and four.
  In particular, the linear bands in the three layers system show a slope very close to the one of graphene,
   suggesting that their Fermi velocities will be similar.

\subsection{Dielectric response} 
The imaginary part of the optical dielectric function is calculated as (see e.g. Ref.\onlinecite{rajeev})

\begin{widetext}
\begin{equation}
\epsilon_2^{ij} (\omega) \propto {1 \over V} \sum_{{\bf k}nn'} <{\bf k}n|p_i|{\bf k}n'><{\bf k}n'|p_j|{\bf k}n>
\times ~ f_{{\bf k}n}(1-f_{{\bf k}n'}) \delta (e_{{\bf k}n'} - e_{{\bf k}n} - \hbar \omega)
\end{equation}
\end{widetext}
where $<{\bf k}n|p_i|{\bf k}n'>$ is the expectation value of the momentum operator between band states $|n>$ and $|n'>$ for states with 
 the crystal momentum ${\bf k}$, i (and j)= x, y or z, and V is the volume of the unit cell of the crystal. In a calculation of graphene, 
  which is a two dimensional object, one has to make a somewhat arbitrary choice of the volume of the C atoms of the two dimensional unit cell.
  We have chosen to use a volume of these C atoms which is the same as the volume of C in graphite. In practice then the calculations were made 
  for a cell with extended c-axis, in order to simulate isolated C layers, and the numerical value of $\epsilon_2^{ij} (\omega)$ was scaled to
  correspond to a volume V which is that of graphite.
 
\begin{figure*}[htbp]
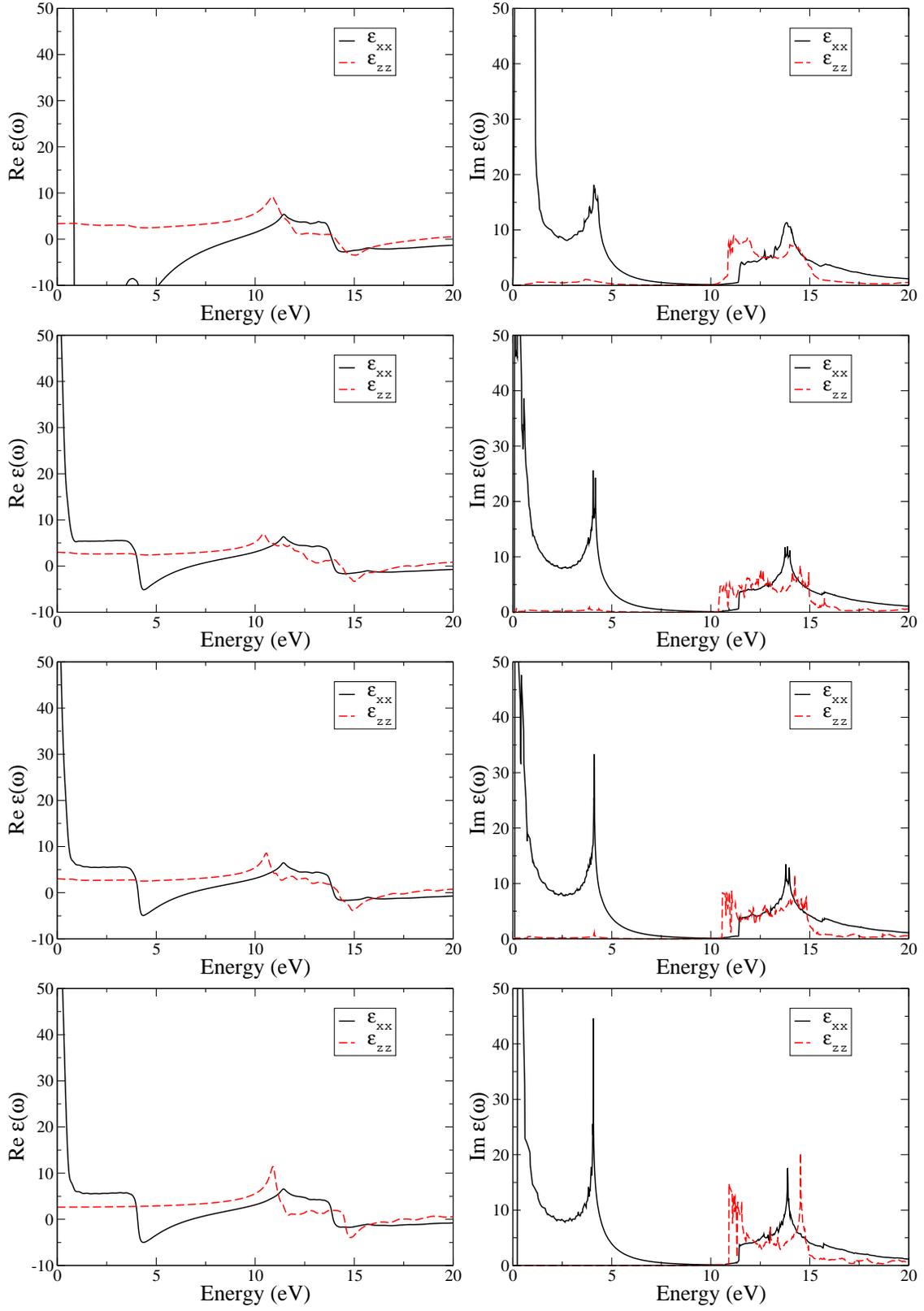

\includegraphics*[angle=0,scale=0.30]{3_re-graphite.ps}
\includegraphics*[angle=0,scale=0.30]{4_im-graphite.ps}
\includegraphics*[angle=0,scale=0.30]{5_re-three.ps}
\includegraphics*[angle=0,scale=0.30]{6_im-three.ps}
\includegraphics*[angle=0,scale=0.30]{7_re-two.ps}
\includegraphics*[angle=0,scale=0.30]{8_im-two.ps}
\includegraphics*[angle=0,scale=0.30]{9_re-one.ps}
\includegraphics*[angle=0,scale=0.30]{10_im-one.ps}
\caption{ {\blue Real (left column) and imaginary (right column) part of the dielectric function of graphite (upper panel), and of three
(second upper panel), two (second lower panel), and one (lower panel) layers of graphene.} }
\label{fig:diel1}
\end{figure*}

  Our computed dielectric functions (real and imaginary parts, albeit without a Drude component) for {\blue graphite and for }one, two, and three graphene layers
are presented respectively in Fig. \ref{fig:diel1}. Results for the x-component of the momentum operator (see Eq. 1) 
($\epsilon_{xx}$) are presented in full (black) lines, while the
results corresponding to the z-component of the momentum operator ($\epsilon_{zz}$) are in dashed (red) lines.
The data in Fig. \ref{fig:diel1} agree rather well with published data for graphite\cite{rajeev}.
In Ref.\onlinecite{rajeev} a comparison between experimental and theoretical data for graphite was made, and it was observed that the agreement was rather
satisfactory. {\blue Also, our results agree well with the one of Marinopoulous\cite{reining}}.
 For graphite\cite{rajeev}, 
 the observed features are mostly due to transitions between $\pi$ and $\pi^{*}$ states (for the $4$ eV peak),
 and to transitions between $\sigma$ and $\sigma^{*}$ states on the high-symmetry line between $\Gamma$
 and $M$. 

The data in Fig. \ref{fig:diel1} suggest that concerning the real and imaginary parts of $\epsilon$, the calculated values are rather independent on the number of graphene layers, i.e. the curves in Fig. \ref{fig:diel1} are essentially independent on the number of C layers. The main effect of the thickness can be found for 
 $\epsilon_{zz}$, where for one single layer of graphene, the transition 
   between $\pi$ and $\pi^{*}$ states are forbidden\cite{rajeev}, so $\epsilon_{zz}$ is exactly zero between
    $0$ and $\sim 10$ eV. For two and three graphene layers, these transitions are not stricly forbidden
     but they remain very weak. 

{\blue However, the imaginary part of the dielectric function of graphite is slighlty different for $\omega$ close to zero from the one of multilayers of
 graphene because the bands near the high-symmetry $K$, which are responsible for the transitions at suck energies, are more splitted in the case
  of graphite. Therefore, the peak for $\omega \rightarrow 0$ of $Im \epsilon_{xx}$ is enlarged, and then $Re \epsilon_{xx}$ comes out differently
   for graphite.}

\subsection{Elastic constants}
The theory of elasticity of three dimensional objects can be cast in a simple equation
\begin{equation}
E(V,\delta) \approx E(V_0,0) + V_0 \sum_i {\tau_i} \varsigma_i \delta_i + {V_0 \over 2} \sum_{ij} {C_{ij}} \varsigma_i \delta_i \varsigma_j \delta_j , \label{eqn:one}
\end{equation}
where $E(V_0,0)$ is the total energy of the undistorted system at volume $V_0$, the sums run over Voigt index 1-6, $\varsigma_i$ takes the value 1 if the Voigt index is 1,2 or 3, and it takes the value 2 if the Voigt index takes the values 4,5 or 6. Furthermore, $\tau_i$ is an element of the stress tensor and $C_{ij}$ is the elastic constant.\cite{fast} For a two dimensional object like graphene the theory of elasticity becomes somewhat modified, as discussed e.g. by Behroozi \cite{behroozi}. Hence, the expression in Eqn.1 is modified to,
\begin{equation}
E(A,\delta) \approx E(A_0,0) + A_0 \sum_i {\tau_i} \delta_i + {A_0 \over 2} \sum_{ij} {C_{ij}} \varsigma_i \delta_i  \varsigma_j \delta_j , \label{eqn:one}
\end{equation}
where $A_0$ is the area of the unit cell. For a three dimensional hexagonal lattice there are 5 elastic constants $C_{11}$, $C_{12}$, $C_{13}$, $C_{33}$, and $C_{55}$, which for the two dimensional hexagonal lattice of graphene reduces to only $C_{11}$ and $C_{12}$. Because the expression in Eqn.2 involves an area instead of a volume in front of the summation, the unit of the two-dimensional elastic constant is different than that of a three dimensional elastic constant, where the unit is Pa. Hence the unit of the two dimensional elastic constant 
is m$\cdot$Pa. We will below report on our calculated elastic constants of graphene in this unit.

\begin{figure*}[htbp]
\includegraphics*[angle=0,scale=0.20]{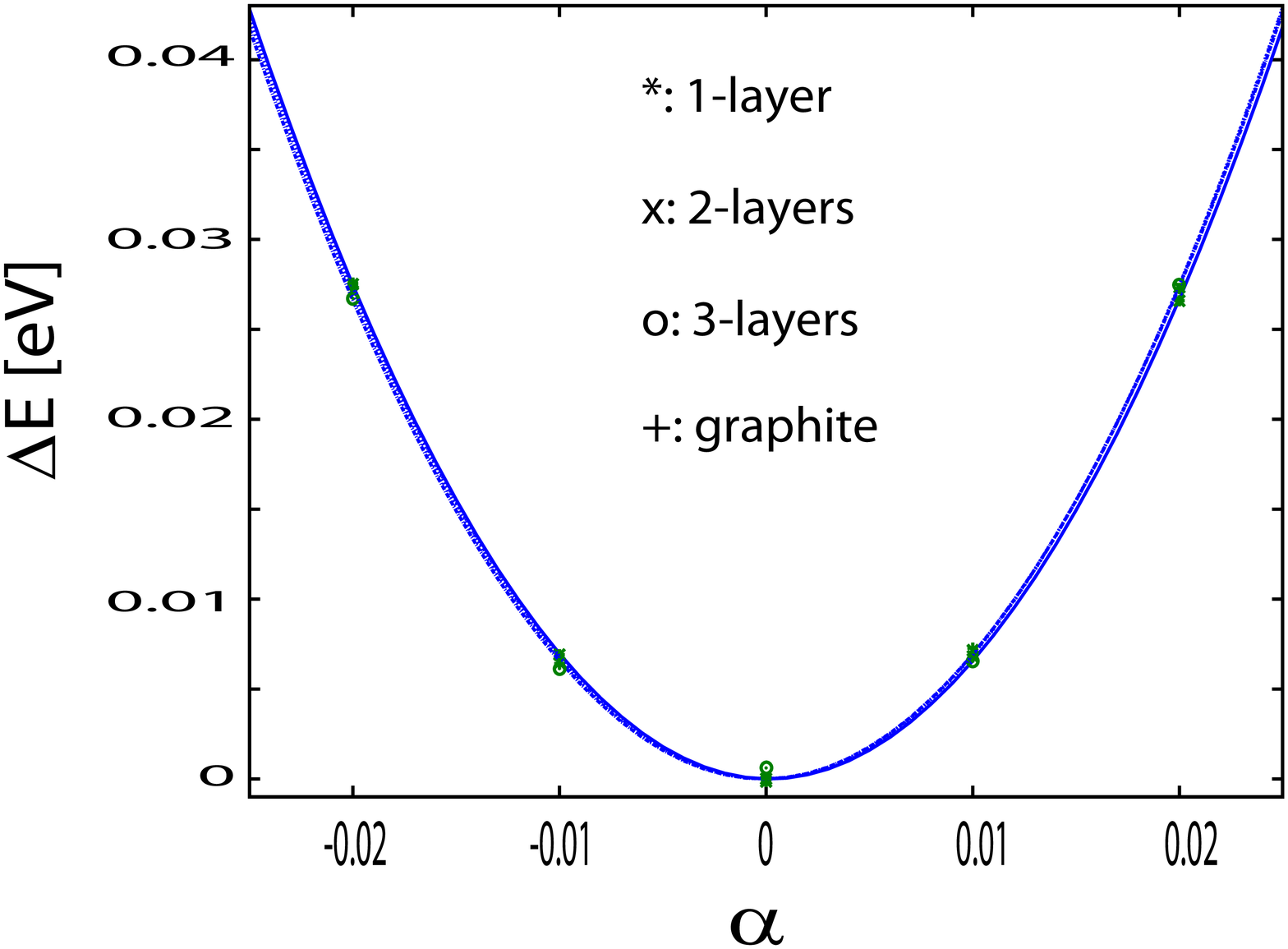}
\caption{The calculated total energy as a function of distortion. $C_{11}+C_{12}$ is calculated using $\alpha_{xx}=\alpha$, $\alpha_{yy}=\alpha$ and with the other elements of the distortion matrix $\alpha_{ij}=0$.}
\label{elast1}
\end{figure*}

\begin{figure*}[htbp]
\includegraphics*[angle=0,scale=0.20]{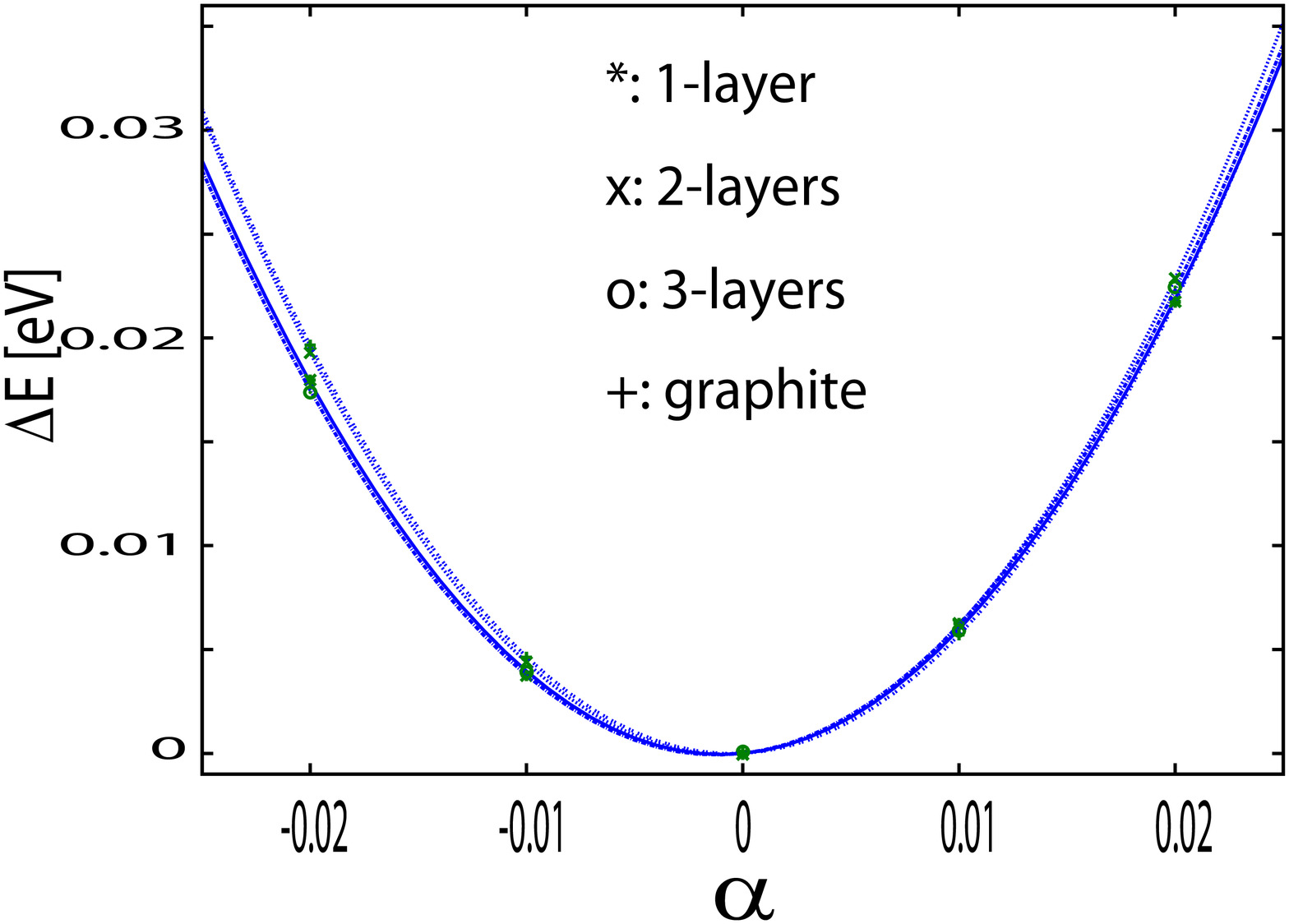}
\caption{The calculated total energy as a function of distortion. $C_{11}-C_{12}$ is calculated using $\alpha_{xx}=\alpha$, $\alpha_{yy}=-\alpha$ and with the other elements of the distortion matrix $\alpha_{ij}=0$.}
\label{elast2}
\end{figure*} 

In Figs. \ref{elast1} and \ref{elast2} we show the calculated total energy versus distortions corresponding to the elastic constants c$_{11}$ and c$_{12}$, respectively. It should be noted that we display the energy per C atom. As is obvious from the two figures the different systems react very similarly to distortion with roughly the same energy cost. Hence, the expansion coefficients for bulk graphite as well as for graphene, bi-layer graphene and tri-layer graphene are all very similar. This is consistent with the fact that the chemical binding which is relevant for these two distortions is governed by the sp$^2$ bonds, which are very similar for the four systems shown in Figs.\ref{elast1} and \ref{elast2} .

For graphite the calculated distortions correspond to a value of c$_{11}$= 1.098 MBar and c$_{12}$=0.154 MBar. Both values reproduce with acceptable accuracy the experimental values, see table.I. This gives credit to the accuracy of the calculations and enables us to trust the elastic constants of graphene, bi-layer graphene and tri-layer graphene. The elastic constants for these systems are also listed in Table.I. Unfortunately we are not aware of experiental data with which to compare these numbers, and hence our theory serves as a prediction. We note however that experimental studies of elasticity of graphene have been published recetly, reporting on the Youngs modulus.\cite{lee} {\red Michel and Verberck \cite{michel} have calculated the elastic constants for graphite and tension coefficients ($c_{ij}\approx 2\gamma_{ij}/c$) for graphene using the Born long wave method to obtain phonon dispersion. Table.I. shows that the elastic constants for graphite reported in \cite{michel} are too large compared to experiment (10\% and 70\% for $c_{11}$ and $c_{12}$, respectively). Because we consistently report smaller elastic constant and tension coefficients compared to \cite{michel} we are confident of the accuracy of the first-principles calculations in this paper.}

\begin{table}
\caption{\label{tab:table1}Elastic constants of graphite [in TPa] and of graphene [in Pa m] for 1, 2 and 3 C layers. The tension coefficients ($\gamma_{11}$, $\gamma_{12}$ and $\gamma_{66}$) for graphene are given in $10^4$ dyn/cm. For completness $c_{66}=(c_{11}-c_{12})/2$ is also given.}
\begin{ruledtabular}
\begin{tabular}{lrrr}
Material&$c_{11}$ ($\gamma_{11}$)&$c_{12}$ ($\gamma_{12}$)&$c_{66}$ ($\gamma_{66}$)\\
\hline
graphite (calc.) & 1.098 & 0.154 & 0.472\\
graphite (calc.\footnote{Ref.\cite{michel}.}) & 1.211 & 0.276 & 0.468 \\
graphite (exp.) & 1.060\footnote{Refs.\cite{blackslee,seldin}.}, 1.109\footnote{Ref.\cite{bosak}.} & 0.180$^b$, 0.139$^c$& 0.442$^b$, 0.485$^c$\\
1-layer (calc.)& 358 (35.8)& 55.0 (5.50)& 152 (15.2)\\
1-layer (calc.$^a$)&  (40.6)&  (9.2)&  (15.7)\\
2-layers (calc.)& 368 (36.8)& 47.3 (4.73)& 160 (16.0)\\
3-layers (calc.)& 358 (35.8)& 54.5 (5.45)& 152 (15.2)\\
\end{tabular}
\end{ruledtabular}
\end{table}

\section{\label{conclusion} Conclusion}
In this paper we have studied theoretically several materials properties when going from one C layer in graphene to two and three graphene layers and on to graphite. The properties we have focused on are the elastic constants, electronic structure (energy bands and density of states), and the dielectric properties. In general we find very similar behaviour for all studied systems. For any of the properties we have looked at the modification due to an increase in the number of graphene layers is within a few percent. The largest effect due to the thickness is found for $\epsilon_{zz}$ which is zero in the energy interval of 0 - $\sim$ 10 eV for monolayer graphene, and non-zero for thicker layers, including graphite. The mililarity in elastic constants, C$_{11}$ and C$_{12}$, for the here studied systems is naturally due to that these constants are determined by the covalent in-plane sp$^2$ hybrids, which are essentially the same and independent on thickness. Our results are in
  agreement with the analysis presented recently by Kopelevich and Esquinazi.\cite{kopelevich}

\section{Acknowledgements}
We acknowledge support from the Swedish Research Council (VR), Swedish Foundation for Strategic Research (SSF), the Swedish National Allocations Committee (SNIC/SNAC), and the G\"oran Gustafsson Stiftelse.
S. L. acknowledges financial support from ANR PNANO Grant ANR-06-NANO-053-02 and ANR Grant ANR-BLAN07-1-186138.

\end{document}